\documentclass[letterpaper]{article} 
\usepackage{aaai2027}  
\usepackage[hyphens]{url}  
\usepackage{graphicx} 
\usepackage{natbib}  
\usepackage{caption} 
\usepackage{algorithm}
\usepackage{algorithmic}

\usepackage{newfloat}
\usepackage{listings}
\DeclareCaptionStyle{ruled}{labelfont=normalfont,labelsep=colon,strut=off} 
\floatstyle{ruled}
\newfloat{listing}{tb}{lst}{}
\floatname{listing}{Listing}

\usepackage{booktabs}

\usepackage{amsmath}
\usepackage{amssymb}
\usepackage{multirow}
\usepackage{subcaption}
\usepackage{tcolorbox}
\tcbuselibrary{breakable}
\usepackage{enumitem}
\usepackage{longtable}

\nocopyright

\def\oursName{DOCG-AS}
\title{From Latent Influence to Language: \\[0.5em] Diffusion-Oriented Content Generation via Audience-Susceptible Features}
\author{
    Jiaying Lei\textsuperscript{\rm 1,\rm 2},
    Shengqi Dang\textsuperscript{\rm 1,\rm 2},
    Runqian Bai\textsuperscript{\rm 1},
    Ziqing Qian\textsuperscript{\rm 1},
    Nan Cao\textsuperscript{\rm 1,\rm 2}\thanks{Corresponding author.}
}
\affiliations{
    \textsuperscript{\rm 1}Tongji University\\
    \textsuperscript{\rm 2}Shanghai Innovation Institute\\
    
}

\begin{document}

\maketitle


\begin{abstract}
The rapid growth of multimodal user-generated content on social media has made information diffusion a critical factor for advertisers and brand marketers. However, manually tailoring content to resonate with specific audiences is labor-intensive and heuristic-driven. While recent generative models offer promising capabilities for automatic content generation, existing approaches for diffusion-oriented content generation still struggle to effectively translate numeric diffusion influence signals into actionable guidance that captures latent audience susceptibility and accounts for heterogeneous audience interests. To address these challenges, we propose \oursName, a three-stage framework for diffusion-oriented content generation. It first performs implicit feature optimization on the realistic content manifold to discover an optimal propagation feature vector. Then, it explicitly decodes this vector using a learnable decoder into interpretable audience-susceptible features described in natural language, providing guidance for content generation. Finally, it leverages multiple sets of audience-susceptible features obtained from different optimization initializations to rewrite the user's input into the final multimodal content. Experiments demonstrate that \oursName~consistently outperforms state-of-the-art baselines in terms of predicted diffusion influence.
\end{abstract}
\section{Introduction}
\label{sec:intro}
The rapid proliferation of social media has led to an unprecedented surge in multimodal user-generated content, intensifying competition for limited audience attention. On these platforms, information propagates through complex user interactions~\cite{kempe2003maximizing}, fundamentally shaping practices in influencer marketing, digital advertising, and brand communication. As a result, diffusion-oriented content---content designed to maximize information spread—has emerged as a critical objective~\cite{berger2012makes}. In practice, creators predominantly rely on heuristic strategies,
manually adjusting linguistic and visual presentation to align
with audience preferences~\cite{barcelos2018watch,dang2026words}. However, such ad hoc processes are labor-intensive and fail to systematically capture the latent factors governing audience engagement.

Recent advances in generative models~\cite{achiam2023gpt,rombach2022high} have enabled high-fidelity text and image synthesis, offering new opportunities for scalable content generation. This has spurred growing interest in persuasive and engagement-driven content generation~\cite{coppolillo2025engagement,matz2024potential}. Nevertheless, existing approaches exhibit notable limitations: most optimize for coarse, audience-agnostic engagement metrics, while others restrict personalization to textual modalities at the trait or persona level~\cite{pillai2025engagement}. Consequently, the problem of multimodal content generation tailored to diffusion within specific audience segments remains insufficiently explored. Recent work~\cite{qian2026designed} takes an initial step by formalizing diffusion-oriented content generation (DOCG) and proposing a reinforcement learning (RL) framework to optimize text and images under the guidance of the predicted influence score. Although promising, this paradigm has limitations: (1) the action space defined in its RL process is predetermined and drawn from an unvalidated marketing framework (i.e., the STEPPS model~\cite{berger2012makes}); (2) the generation process incurs substantial computational overhead due to the heavy estimation required for iterative generation and evaluation. These limitations underscore the need for a more flexible and efficient approach to tackling DOCG.

Despite recent advances in generative modeling, diffusion-oriented content generation remains fundamentally challenging due to three core issues.
\textbf{(1) Elusive latent susceptibility.} Information diffusion within a target audience is governed by higher-order, implicit factors that are difficult to disentangle from observational data. For example, a movie post may spread because a sarcastic question is paired with a high-contrast still.
\textbf{(2) Non-actionable diffusion signals.} Although computational information diffusion models provide estimates of propagation (e.g., influence score), such a score may show whether a post is more likely to spread than another, but not how to adjust its content (e.g., emotional tone and headline) for improved diffusion performance.
\textbf{(3) Heterogeneous and context-dependent preferences.} Beyond shared coarse preferences (e.g., movie enthusiasm), users exhibit significant intra-group divergence along secondary dimensions (e.g., avant-garde vs. classical tastes). This heterogeneity induces an optimization landscape with multiple local maxima, which hinders content optimization.
   
To address these challenges, we propose a novel three-stage framework for diffusion-oriented content generation via audience-susceptible feature discovery, which takes a target audience group and source content as inputs. The first module, \textbf{implicit feature optimization}, identifies the propagation-optimal feature vector $\boldsymbol{e}^*$ (i.e., the latent representation that maximizes diffusion within the audience) using an influence indicator trained on large-scale user interaction data.  
The second module, \textbf{explicit feature decoding}, maps $\boldsymbol{e}^*$ into structured natural-language descriptions along eight well-established, reliable dimensions (theme, structure, rhetoric, and emotion for both textual and visual modalities, as illustrated in Figure~\ref{fig:dimensions}) via a learnable LLM-based decoder, thereby bridging continuous latent representations with discrete, controllable semantics.
The third module, \textbf{audience-susceptible content generation}, uses multiple decoded feature sets obtained from diverse optimization initializations to capture audience heterogeneity and rewrites the input into engaging multimodal content using off-the-shelf generative models. Experiments and a user study demonstrate substantial improvements
in predicted diffusion influence and human preference while preserving the source content's core intent. Our contributions are summarized as follows:
\begin{itemize}
\item We propose a novel three-stage framework for multimodal diffusion-oriented content generation via audience-susceptible feature discovery.
\item We introduce an implicit feature optimization method that identifies propagation-optimal features on the realistic content manifold, jointly improving diffusion influence while maintaining semantic plausibility.
\item We develop an explicit feature decoding mechanism that maps continuous optimal features to structured, interpretable natural-language representations, effectively bridging latent optimization and the discrete semantic space of LLMs.
\item We design an audience-susceptible content generation strategy that leverages diverse decoded feature sets to model audience heterogeneity and enables controllable multimodal content rewriting.
\end{itemize}

\section{Related Work}
\label{sec:related}

\begin{figure*}[t]
    \centering\includegraphics[width=1.0\linewidth]{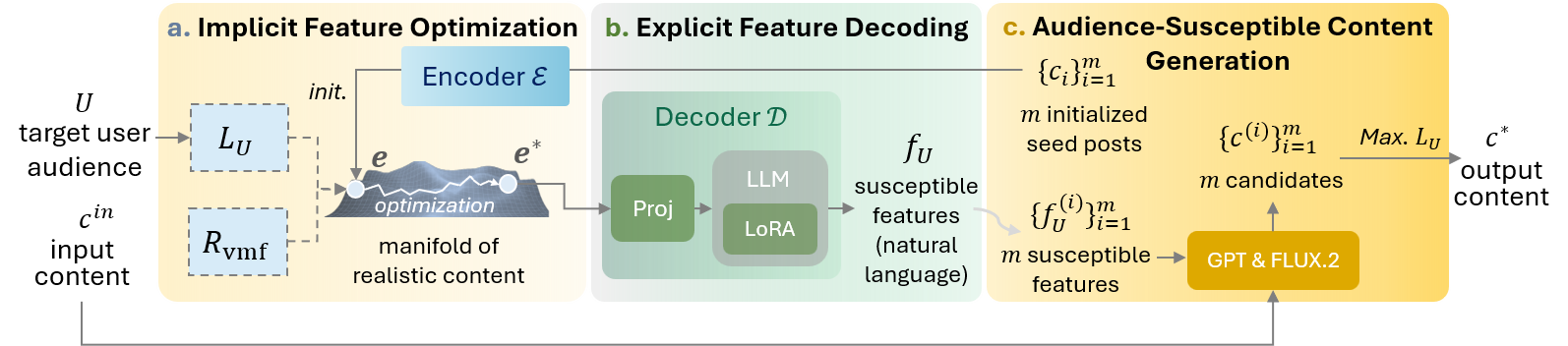}
    \caption{Pipeline of the proposed framework for generating engaging multimodal content $c^*$ tailored to a target audience group $U$, conditioned on input content $c^{\mathrm{in}}$.}
    \label{fig:pipeline}
\end{figure*}

We review related work from two perspectives: (1) content-level factors for promoting information diffusion and (2) diffusion-oriented content generation.

\subsection{Content-Level Factors for Promoting Diffusion}

A large body of research shows that multiple factors influence the effectiveness of information diffusion. For the textual modality, information value and linguistic style affect how widely a message spreads~\cite{dvzanko2025linguistic}, while content organization and specific wording also play important roles~\cite{gligoric2019causal,borghouts2024wording}. Engagement-oriented text rewriting further demonstrates that changes in linguistic style can significantly alter a post's predicted engagement~\cite{pillai2025engagement}. 
For the visual modality, the mere presence of images or videos can increase retweeting behavior~\cite{xie2022identifying}, and more fine-grained automated visual analysis further reveals that objects, scenes, people, facial or bodily cues, and various visual attributes all influence social media effects~\cite{peng2024automated}. Beyond depicted content, image color, visual quality, and the volume, variety, and dynamism of visual elements are also closely related to user engagement, viewers' emotional responses, and subsequent behavioral responses~\cite{yu2021color,kanuri2024standing,philp2022predicting,chan2023more}. 

Building on these findings, we construct eight content dimensions for the DOCG task.
Compared with the diffusion-oriented editing dimensions defined in Designed2Spread~\cite{qian2026designed}, our dimensions more clearly separate content, organization, expression, and audience response across both text and image modalities.


\subsection{Diffusion-Oriented Content Generation}

Recent work has explored automatically generating content optimized for social media spread. Early approaches focus on maximizing generic engagement signals. For instance, some methods fine-tune large language models against simulated diffusion feedback~\cite{coppolillo2025engagement} or align generation with popularity indicators such as the number of likes and retweets~\cite{yu2024repalm}, while others inject network structure into prompts to amplify post influence~\cite{zhao2025amplifying}. 
A step further, personality-aligned text generation has been shown to elicit stronger persuasive responses at the trait level~\cite{matz2024potential}, yet it targets coarse-grained audience segments rather than specific groups.

The most recent work~\cite{qian2026designed} explicitly tackles multimodal diffusion-oriented content generation. It proposes a reinforcement learning framework that rewrites text or images using a predefined set of editing actions, guided by a content-level influence indicator. However, this paradigm suffers from three key limitations. First, its hand-crafted action primitives cannot discover latent susceptibility patterns beyond human specification. Second, it optimizes against an opaque scalar reward that offers no interpretable guidance for generation. Third, its deterministic policy overlooks the inherent heterogeneity of audience preferences.

In contrast, our approach addresses these limitations directly. We replace hand-crafted actions with implicit feature optimization on the realistic content manifold to uncover latent susceptibility patterns. We decode the optimized features into structured natural-language descriptions, providing interpretable guidance. Finally, we adopt multi-start optimization to capture audience heterogeneity, all within a unified multimodal generation pipeline.

\section{Method}
\label{sec:method}

\begin{figure*}[t]
    \centering\includegraphics[width=1.0\linewidth]{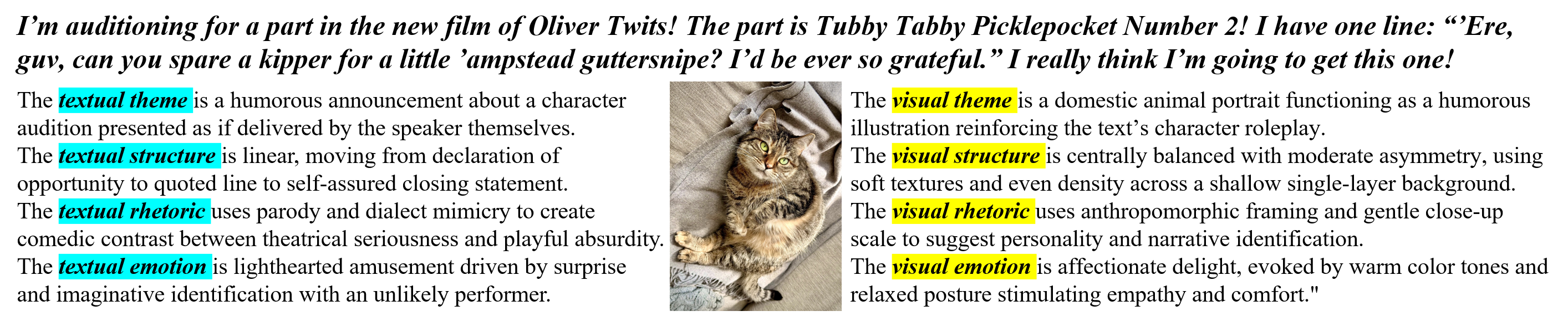}
    \caption{An illustrative example of a social media post and its corresponding natural-language descriptions across the eight defined semantic dimensions.}
    \label{fig:dimensions}
\end{figure*}

In this section, we present the proposed three-stage framework for generating engaging multimodal content $c^*$ tailored to a target audience group $U$, conditioned on input content $c^{\mathrm{in}}$. As illustrated in Figure~\ref{fig:pipeline}, the framework takes as input (1) a target audience group $U$ and (2) source content $c^{\mathrm{in}}$, and proceeds through three modules:
(1) \textit{Implicit Feature Optimization}, which identifies a propagation-optimal feature vector on the realistic content
manifold given $U$;
(2) \textit{Explicit Feature Decoding}, which converts the optimized feature into audience-susceptible features, namely structured and interpretable natural-language descriptions spanning multiple semantic dimensions; and
(3) \textit{Audience-Susceptible Content Generation}, which leverages the decoded features to rewrite $c^{\mathrm{in}}$ into engaging multimodal content $c^*$.

\subsection{Implicit Feature Optimization}
\label{sec:implicit_opt}

In this step, we aim to identify the latent feature representation (denoted as $\boldsymbol{e}^*$) that maximizes diffusion influence within the target audience group $U$. We base this on an influence indicator $L_U(\boldsymbol{e})$ adapted from~\cite{qian2026designed} (we slightly modify the formulation to fit our setting; details are provided in Supplementary Material). Specifically, $L_U(\boldsymbol{e})$ is trained on real Twitter/X data that capture users' diffusion behaviors through observed retweets. It estimates the influence of a message $c$, encoded as $\boldsymbol{e} = \mathcal{E}(c)$, by averaging the predicted pairwise propagation probabilities $p_{ij}$ among users in $U$:
\begin{equation}
   L_U(\boldsymbol{e}) = \frac{1}{|U|(|U|-1)}\sum_{\substack{i,j \in U \\ i \neq j}} p_{ij}(\boldsymbol{e})
\end{equation}
Next, we derive the optimal feature $\boldsymbol{e}^*$ by maximizing the influence indicator $L_U(\boldsymbol{e})$ under a realism constraint that discourages out-of-distribution feature drift, formulated as:
\begin{equation}
\boldsymbol{e}^* = \arg\max_{\boldsymbol{e}}\,\,
L_U(\boldsymbol{e}) + \lambda \mathcal{R}_{\mathrm{vmf}}(\boldsymbol{e})
\label{eq:prototype}
\end{equation}
where $\mathcal{R}_{\mathrm{vmf}}(\boldsymbol{e})$ is a manifold regularizer that constrains $\boldsymbol{e}$ to lie in high-density regions of the distribution of real-world content features, thereby promoting semantically valid and realizable representations, and $\lambda > 0$ is a regularization weight that balances the two terms.
Specifically, $\mathcal{R}_{\mathrm{vmf}}(\boldsymbol{e})$ is defined via a von Mises-Fisher (vMF) kernel density estimator~\cite{banerjee2005clustering} over a set of $K$ reference multimodal posts $\{c_k\}_{k=1}^{K}$ sampled from a real-world dataset associated with the target audience $U$:
\begin{equation}
  \mathcal{R}_{\mathrm{vmf}}(\boldsymbol{e})
  \;=\;
  \tau \cdot \log \sum_{k=1}^{K} \exp\!\left( \frac{\boldsymbol{e}^{\top} \mathcal{E}(c_k)}{\tau} \right)
  \label{eq:vmf}
\end{equation}
where $\tau>0$ controls the concentration of the estimator,
and $\mathcal{E}(\cdot)$ is a pre-trained multimodal encoder (Qwen3-VL-Embedding-8B~\cite{li2026qwen3} in our implementation) that maps raw content to normalized feature representations. We further $\ell_2$-normalize $\boldsymbol{e}$ after each gradient update to keep it on the unit hypersphere. Maximizing this regularizer encourages $\boldsymbol{e}$ to align with high-density regions of the empirical feature distribution, thereby discouraging drift into out-of-distribution areas and ensuring semantic plausibility.

\subsection{Explicit Feature Decoding}
\label{sec:explicit_decode}

In this step, we design a decoder $\mathcal{D}$ based on a large language model (LLM) that translates the optimized feature $\boldsymbol{e}^*$ into explicit audience-susceptible features $f_U$, expressed as structured natural-language descriptions, enabling interpretability and facilitating downstream content generation. This design is motivated by two considerations: (1) LLMs encode rich prior knowledge of linguistic and rhetorical patterns, enabling the generation of coherent and semantically grounded descriptions; and (2) expressing the decoded features in natural language makes them directly usable by downstream LLM-based generation modules, eliminating the need for additional interfaces or adaptation layers.

Specifically, based on our literature review, we define four semantic dimensions for both textual and visual modalities: theme, structure, rhetoric, and emotion. This yields eight modality-specific dimensions: \{\underline{\textit{Textual Theme}, \textit{Visual}} \underline{\textit{Theme}, \textit{Textual Structure}, \textit{Visual Structure}, \textit{Textual Rhetoric},} \underline{\textit{Visual Rhetoric}, \textit{Textual Emotion}, \textit{Visual Emotion}}\}, as illustrated in Figure~\ref{fig:dimensions}. We describe the decoder's architecture and training paradigm in detail below.

Formally, the decoding process is defined as:
\begin{equation}
  f_U \;=\; \mathcal{D}(\boldsymbol{e}^*),
  \label{eq:decode_def}
\end{equation}
where $\mathcal{D}$ translates the numerical feature vector $\boldsymbol{e}^*$ into natural-language descriptions $f_U$ in two steps.

First, we project $\boldsymbol{e}^*$ into the token embedding space of the LLM (Qwen3-14B~\cite{yang2025qwen3} in our implementation) via a cross-attention projector:
\begin{equation}
    \boldsymbol{p} \;=\; \mathrm{Proj}(\boldsymbol{e}^*;\, \boldsymbol{W}_p),
\end{equation}
where $\boldsymbol{p}$ is a sequence of vectors representing the projection of $\boldsymbol{e}^*$ in the LLM's token embedding space, also referred to as soft prompt tokens~\cite{lester2021power}, and $\boldsymbol{W}_p$ denotes the learnable projection parameters. The soft prompt tokens can be directly fed into the LLM as input, enabling the LLM to interpret the semantic meaning of $\boldsymbol{e}^*$ through autoregressive generation.

Next, we adopt Low-Rank Adaptation (LoRA)~\cite{hu2022lora} to fine-tune the LLM in a parameter-efficient manner for more contextually grounded and meaningful feature decoding in the social media domain. The decoding process is then defined as:
\begin{equation}
    \mathcal{D}(\boldsymbol{e}^*) \;=\; \mathrm{LLM}^{\mathcal{D}}\big(\langle \boldsymbol{p}_{\mathrm{ins}}, \boldsymbol{p} \rangle;\, \boldsymbol{W}_l\big),
\label{eq:decode_process}
\end{equation}
where $\mathrm{LLM}^{\mathcal{D}}(\cdot)$ denotes the autoregressive decoding process of the LLM, which takes as input the token embeddings $\boldsymbol{p}_{\mathrm{ins}}$ of a predefined instruction (e.g., ``describe the textual and visual theme, structure, rhetoric, and emotion of this post'') concatenated with $\boldsymbol{p}$, and outputs the decoded features as structured natural-language descriptions. $\boldsymbol{W}_l$ denotes the LoRA parameters, and the instruction serves as a task-specific prompt to guide the decoding process.
    
To train the decoder $\mathcal{D}$, we first construct a large-scale training corpus $\mathcal{S} = \{(c_i, f_i)\}_{i=1}^{N}$, where each empirical post $c_i$ is annotated by GPT-5~\cite{openai2025gpt5} with target natural-language descriptions $f_i$ across the
applicable textual and visual dimensions.
During training, we feed $\boldsymbol{e}_i = \mathcal{E}(c_i)$ into the decoder, and optimize $\mathcal{D}$ using the standard token-level cross-entropy loss for autoregressive language modeling:
\begin{equation}
    \mathcal{L}_{\mathrm{dec}} \;=\; -\frac{1}{N} \sum_{i=1}^{N} \sum_{j=1}^{M_i} \log P_{\mathcal{D}}\big(y_{i,j} \mid y_{i,<j}, \boldsymbol{e}_i\big),
    \label{eq:decoder_loss}
\end{equation}
where each target feature $f_i$ is tokenized into a sequence of $M_i$ tokens $\{y_{i,1}, y_{i,2}, \dots, y_{i,M_i}\}$, $P_{\mathcal{D}}(\cdot \mid \cdot)$ denotes the next-token distribution induced by the decoder, and $y_{i,<j}$ represents the previously generated tokens. $\mathcal{E}(\cdot)$ is the same pre-trained multimodal encoder introduced above.

We jointly optimize the projection parameters $\boldsymbol{W}_p$ and the LoRA adapters $\boldsymbol{W}_l$ in an end-to-end manner, aligning numerical features with the LLM's semantic space while improving linguistic quality and structural consistency. To further enhance robustness, we inject Gaussian noise into $\boldsymbol{e}_i$ during training, encouraging the decoder to remain stable under feature perturbations.

\subsection{Audience-Susceptible Content Generation}
\label{sec:generation}
Finally, we generate the optimal multimodal content $c^*$ that maximizes diffusion influence within the target audience, conditioned on the decoded audience-susceptible features $f_U$ and the source content $c^{\mathrm{in}}$. Notably, the optimization objective in Eq.~\eqref{eq:prototype} is inherently non-convex, so different initializations may converge to distinct local optima $\boldsymbol{e}^*$. This suggests that the audience $U$ is characterized by multiple coexisting interest modes rather than a single homogeneous preference. Capturing such diversity is essential for accurately modeling audience heterogeneity and for generating highly effective, audience-susceptible content.

To account for this, instead of relying on a single solution, we consider multiple audience-susceptible feature optima $\{\boldsymbol{e}_1^*, \boldsymbol{e}_2^*, \dots, \boldsymbol{e}_m^*\}$ obtained by solving Eq.~\eqref{eq:prototype} from $m$ diverse initialization points $\{\mathcal{E}(c_1), \mathcal{E}(c_2), \dots, \mathcal{E}(c_m)\}$. 
Here, $\{c_1, c_2, \dots, c_m\}$ are seed posts selected from the dataset via farthest-point traversal~\cite{gonzalez1985clustering} to ensure coverage and diversity. 
Based on these feature optima, we generate $m$ candidate multimodal posts $\{c^{(i)}\}_{i=1}^{m}$, with each $c^{(i)}$ containing textual content $c_t^{(i)}$ and visual content $c_v^{(i)}$:
\begin{equation}
  \langle c_t^{(i)}, p^{(i)} \rangle
  =
  \mathcal{G}_{\mathrm{text}}
  \big(c^{\mathrm{in}},\, \mathcal{D}(\boldsymbol{e}_i^*)\big),
\end{equation}
\begin{equation}
  c_v^{(i)}
  =
  \mathcal{G}_{\mathrm{img}}
  \big(p^{(i)}\big),
\end{equation}
where $\mathcal{G}_{\mathrm{text}}(\cdot)$ denotes a large language model that generates textual content and an image-generation prompt $p^{(i)}$, $\mathcal{G}_{\mathrm{img}}(\cdot)$ denotes a text-to-image generator that synthesizes visual content (GPT-4o~\cite{hurst2024gpt} and FLUX.2~\cite{flux-2-2025}, respectively, in our implementation). $\mathcal{D}(\cdot)$ is the decoder introduced in Eq.~\eqref{eq:decode_process}. 
Specifically, $\mathcal{G}_{\mathrm{text}}$ uses $\mathcal{D}(\boldsymbol{e}_i^*)$ to rewrite $c^{\mathrm{in}}$ according to the decoded textual theme, structure, rhetoric, and emotion, while generating $p^{(i)}$ based on $c^{\mathrm{in}}$ and the corresponding decoded visual features along the same four dimensions.
The final output is chosen as the candidate with the highest predicted diffusion influence:
\begin{equation}
c^*
\;=\;
\underset{c^{(i)},\, i \in \{1, \dots, m\}}
{\operatorname{arg\,max}}\;
L_U\big(\mathcal{E}(c^{(i)})\big).
\end{equation}

\section{Experiments}
\label{sec:experiments}

\begin{figure*}[t]
    \centering\includegraphics[width=1.0\linewidth]{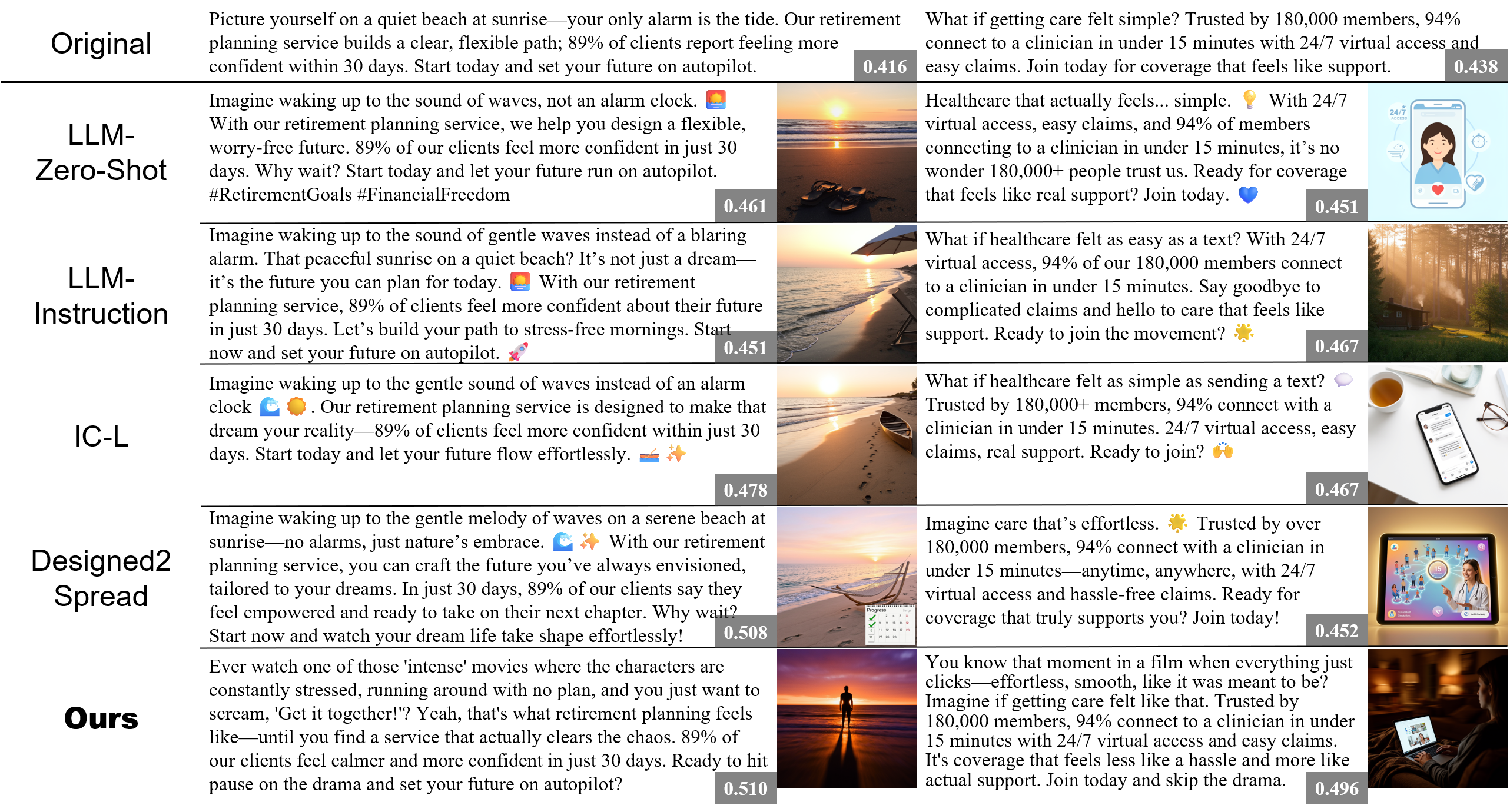}
\caption{Qualitative comparison with baselines on Movie. Each multimodal post is annotated with its predicted influence score.}
    \label{fig:qual-comp}
\end{figure*}

\begin{table}[t]
\centering

\small
\setlength{\tabcolsep}{3pt}

\resizebox{\columnwidth}{!}{%
\begin{tabular}{@{}clcc@{}}
\toprule
\textbf{Dataset}
& \textbf{Method}
& \textbf{Diffusion Gain (\%)} $\uparrow$
& \textbf{Consistency} $\uparrow$ \\
\midrule

\multirow{5}{*}{Movie}
& LLM-Zero-shot
& 0.55 $\pm$ 0.60 
& \textbf{0.8722 $\pm$ 0.0031} \\

& LLM-Instruction
& 1.15 $\pm$ 0.61 
& 0.8680 $\pm$ 0.0030 \\

& IC-L
& 1.30 $\pm$ 0.13 
& 0.8654 $\pm$ 0.0061 \\

& Designed2Spread
& 4.49 $\pm$ 0.13 
& 0.8648 $\pm$ 0.0010 \\

& \textbf{Ours}
& \textbf{16.51 $\pm$ 0.41}
& 0.7783 $\pm$ 0.0257 \\

\midrule

\multirow{5}{*}{SpaceX}
& LLM-Zero-shot
& 2.72 $\pm$ 2.01 
& 0.8622 $\pm$ 0.0027 \\

& LLM-Instruction
& 5.44 $\pm$ 2.13 
& 0.8546 $\pm$ 0.0036 \\

& IC-L
& 6.91 $\pm$ 2.75 
& \textbf{0.8637 $\pm$ 0.0042} \\

& Designed2Spread
& 12.01 $\pm$ 1.72 
& 0.8580 $\pm$ 0.0037 \\

& \textbf{Ours}
& \textbf{40.70 $\pm$ 2.01}
& 0.7756 $\pm$ 0.0052 \\

\bottomrule
\end{tabular}%
}
\caption{Comparison results on Movie and SpaceX (mean $\pm$ standard deviation
over three runs; best results in bold).}
\label{tab:main_results}
\end{table}

We evaluate the proposed framework through quantitative and qualitative comparisons with baselines, a user study, and ablation studies.

\subsection{Experimental Setups}
\label{sec:setups}

\paragraph{Implementation Details.}
We implement the three modules of our framework as follows. For feature optimization, we set the regularization weight $\lambda=5.0$ and the temperature $\tau=0.05$, and optimize each candidate via gradient ascent using the AdamW
optimizer with a learning rate of $0.5$ for $300$ steps. For feature decoding, the feature decoder $\mathcal{D}$ comprises a cross-attention projector that maps the optimized feature $\boldsymbol{e}^*$ into $16$ soft prompt tokens, which are prepended to a Qwen3-14B backbone quantized to 4-bit NF4. The projector parameters $\boldsymbol{W}_p$ and the LoRA adapters $\boldsymbol{W}_l$ with rank $16$ and $\alpha=32$ are jointly fine-tuned for up to $30$ epochs, with Gaussian noise of standard deviation $\sigma=0.005$ injected into $\boldsymbol{e}_i$ during training to enhance robustness. For content generation, we adopt a multi-start strategy with $m=6$ for Movie and $m=10$ for SpaceX. The $m$ initialization points are selected from the seed tweets via farthest-point traversal, with the seed post having the highest $L_U$ selected first. 

\paragraph{Baselines.}
We compare our approach with four baselines spanning three categories, distinguished by their level of audience awareness and whether training is required (see Supplementary Material): 
(1) \emph{Audience-agnostic}. \textbf{LLM-Zero-shot} directly prompts an LLM to rewrite the input content without any audience-specific guidance. \textbf{LLM-Instruction} augments the prompt with manually specified dimension-level instructions, but still lacks explicit modeling of the target audience.
(2) \emph{Audience-aware, training-free}. \textbf{In-Context Learning (IC-L)}~\cite{brown2020language} retrieves, for each audience group, high- and low-influence messages (measured by $L_U$) as positive and negative exemplars, respectively, to guide generation via few-shot prompting.
(3) \emph{Audience-aware, training-based}. \textbf{Designed2Spread}~\cite{qian2026designed} employs a single-agent rewriting framework, leveraging STEPPS~\cite{berger2012makes} and a predefined visual-empirical feature action space to iteratively optimize content for diffusion.
For a fair comparison, all methods use the same text-generation
LLM and image generator, with identical inference settings,
to produce the final multimodal content.

\paragraph{Datasets.}
We construct two Twitter/X datasets, each representing a distinct interest-driven audience group. The \textbf{Movie} dataset focuses on movie-related discussions and contains $124{,}336$ users and $2{,}910{,}600$ tweets ($1{,}699{,}675$ with images). The \textbf{SpaceX} dataset targets SpaceX-related topics and comprises $89{,}663$ users and $2{,}217{,}930$ tweets ($620{,}025$ with images). Following prior work~\cite{qian2026designed}, data collection proceeds in three steps: we first retrieve $5{,}000$ seed tweets for each topic (posted between July 2024 and December
2025) via keyword-based search, then construct the target audience group by identifying users who directly engage with these tweets, and finally aggregate the historical posts of these users to form the dataset.

For feature optimization, the vMF kernel density estimator in Eq.~\eqref{eq:vmf} uses all seed tweets as the reference set.
For feature decoding, we train a separate decoder for each dataset on its own annotated corpus, comprising both seed tweets and additional relevant content, with a $3{:}7$ ratio of text-only to multimodal posts and a total of $75{,}000$ annotated posts across the two datasets (see Supplementary Material).
For content generation, we use GPT-5 to produce $200$ short advertisements for non-tangible products and services as evaluation inputs. Among them, the $160$ samples with the highest predicted influence $L_U$ are used to train Designed2Spread, while the remaining $40$ form a shared test set for all methods.

\paragraph{Metrics.}
Following prior work~\cite{qian2026designed}, we evaluate the generated content $c^*$ using two complementary metrics. (1) \emph{Diffusion Gain}: the relative percentage increase in the predicted diffusion influence score of the generated content $c^*$ over the input content $c^{\mathrm{in}}$, measuring improvement in expected propagation. (2) \emph{Consistency}: the cosine similarity between the feature representations of $c^*$ and $c^{\mathrm{in}}$, quantifying the degree to which the generated content remains semantically consistent with the input.

\subsection{Comparison Results}
\label{sec:generation_eval}

\paragraph{Quantitative Comparison.}
Table~\ref{tab:main_results} summarizes the performance of methods in terms of \emph{Diffusion Gain} and \emph{Consistency}. Our approach consistently achieves the highest Diffusion Gain across both datasets, with a substantial margin over all baselines. On the \textbf{Movie} dataset, it attains a gain of $16.51\%$, substantially outperforming the strongest baseline. The advantage is even more pronounced on \textbf{SpaceX}, where our method reaches $40.70\%$, compared to $12.01\%$ for Designed2Spread, $6.91\%$ for IC-L, and $5.44\%$ for LLM-Instruction.
In terms of Consistency, our method achieves $0.7783$ on Movie and $0.7756$ on SpaceX, slightly lower than more conservative baselines (e.g., LLM-Zero-shot at $0.8722$). This reflects an inherent trade-off: maximizing audience resonance requires non-trivial modifications to the original content, which reduces consistency with the input. Importantly, these scores remain reasonably high, while the qualitative comparison further confirms that our method preserves the core message and semantics despite adapting the textual and visual presentation. The same pattern is observed across additional LLM backbones (details are provided in Supplementary Material).

\paragraph{Qualitative Comparison.}
Figure~\ref{fig:qual-comp} illustrates the effectiveness of our framework in translating latent audience-susceptible features into coherent multimodal content. In contrast to standard baselines (e.g., LLM-Zero-shot, IC-L), which tend to produce generic text and imagery closely aligned with the literal prompt, our method adaptively reshapes both narrative and visual semantics to enhance audience resonance. For example, in the retirement planning scenario (left), baseline methods generate predictable ``sunny beach'' visuals accompanied by conventional promotional language. By contrast, our approach introduces a high-arousal narrative hook (e.g., framing life as a ``stressed movie character'') together with a dramatic, high-contrast visual composition, effectively positioning the service as a compelling resolution to uncertainty. Similarly, for the healthcare prompt (right), while Designed2Spread relies on abstract, impersonal digital motifs, our framework produces content grounded in relatable, empathetic contexts---combining conversational text with warm, intimate domestic imagery. These results demonstrate that our diffusion-oriented framework
goes beyond surface-level prompt alignment by operationalizing
latent audience-susceptible patterns to generate semantically
coherent and engaging content.

\paragraph{User Study.}
To complement the above metrics with human perception, we conducted a user study on the Prolific platform~\cite{prolific} to assess whether our generated content is more likely to be shared in realistic settings. We recruited $50$ participants who actively engage with movie-related content on Twitter/X. Each participant evaluated about $100$ pairs of multimodal posts, where each pair consisted of one post generated by our method and one by the strongest baseline, Designed2Spread, presented in randomized order.
For each pair, participants selected the version they would be more willing to repost, comment on, or share. Following~\cite{qian2026designed}, we report the mean participant-level \textbf{\textit{Repost Preference Rate} (RPR)}, defined for each participant as the proportion of comparisons in which our method is preferred over the baseline. Our approach achieves a mean participant-level RPR of $\mathbf{64.27\%}$, significantly above chance under a one-sided one-sample $t$-test against $0.5$ ($p=6.94\times10^{-7}$). These results further demonstrate that our framework produces more engaging multimodal content from the perspective of real users (see Supplementary Material).

\begin{table}[t]
\centering

\small
\setlength{\tabcolsep}{3pt}

\resizebox{\columnwidth}{!}{%
\begin{tabular}{@{}clcc@{}}
\toprule
\textbf{Dataset}
& \textbf{Method}
& \textbf{Diffusion Gain (\%)} $\uparrow$
& \textbf{Consistency} $\uparrow$ \\
\midrule

\multirow{5}{*}{Movie}
& w/o $\mathcal{R}_{\mathrm{vmf}}$
& 14.92 $\pm$ 0.68
& 0.7771 $\pm$ 0.0147 \\

& w/o Feature Opt.
& 16.07 $\pm$ 0.40
& 0.7697 $\pm$ 0.0107 \\

& w/o Feature Decoding
& 7.96 $\pm$ 0.44
& \textbf{0.8663 $\pm$ 0.0016} \\

& w/o Multi-Start
& 11.57 $\pm$ 0.34
& 0.8170 $\pm$ 0.0102 \\

& \textbf{Ours}
& \textbf{16.51 $\pm$ 0.41}
& 0.7783 $\pm$ 0.0257 \\

\midrule

\multirow{5}{*}{SpaceX}
& w/o $\mathcal{R}_{\mathrm{vmf}}$
& 33.83 $\pm$ 1.47
& 0.8115 $\pm$ 0.0102 \\

& w/o Feature Opt.
& 37.90 $\pm$ 0.51
& 0.7914 $\pm$ 0.0074 \\

& w/o Feature Decoding
& 23.84 $\pm$ 0.40
& \textbf{0.8520 $\pm$ 0.0027} \\

& w/o Multi-Start
& 34.00 $\pm$ 0.25
& 0.8081 $\pm$ 0.0007 \\

& \textbf{Ours}
& \textbf{40.70 $\pm$ 2.01}
& 0.7756 $\pm$ 0.0052 \\

\bottomrule
\end{tabular}%
}
\caption{Ablation results on Movie and SpaceX (mean $\pm$ standard deviation
over three runs; best results in bold).}
\label{tab:ablation}
\end{table}

\subsection{Ablation Study}

\begin{figure}[!htb]
\centering
\includegraphics[width=1\linewidth]{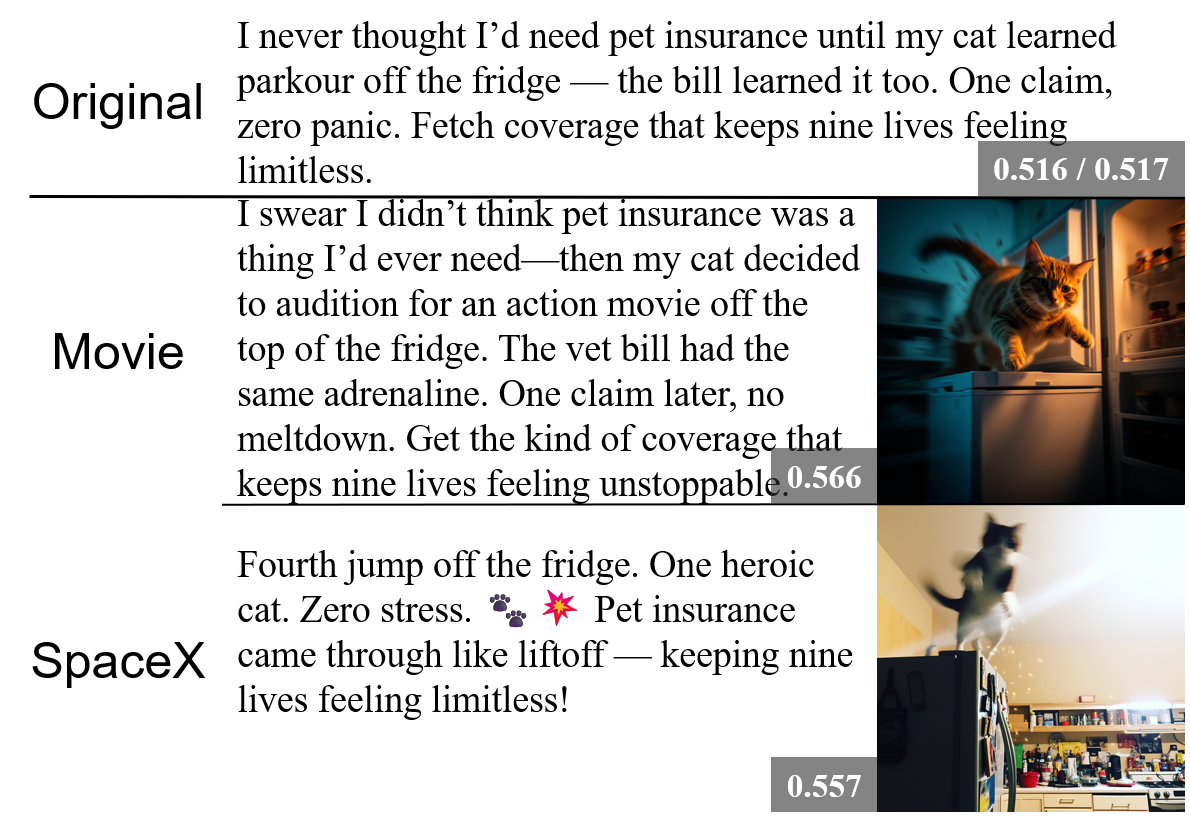}
\caption{Audience-specific generations for Movie and SpaceX from the same source content. Each generated multimodal post is annotated with its predicted influence score for the corresponding dataset, while the source post is annotated with both scores in Movie/SpaceX order.}
\label{fig:qual-pair}
\end{figure}

We evaluate the vMF regularizer in Feature Optimization and the contribution of the three main modules.

\paragraph{w/o vMF Regularization ($\mathcal{R}_{\mathrm{vmf}}$).}
The $\mathcal{R}_{\mathrm{vmf}}$ term constrains the optimized feature $\boldsymbol{e}^*$ to high-density regions of the realistic content manifold. Removing it reduces Diffusion Gain on both datasets, including a drop from $40.70\%$ to $33.83\%$ on SpaceX, as the optimization may drift toward out-of-distribution regions. Although Consistency increases to $0.8115$ on SpaceX, the decoded features are less effective at increasing predicted
diffusion influence. This result demonstrates the importance of vMF regularization for obtaining influential and plausible features.

\paragraph{w/o Feature Optimization.}
We remove Feature Optimization by directly decoding each initial representation $\mathcal{E}(c_i)$ without solving Eq.~\eqref{eq:prototype}. Diffusion Gain decreases from $16.51\%$ to $16.07\%$ on Movie and from $40.70\%$ to $37.90\%$ on SpaceX. Consistency changes only slightly, decreasing from $0.7783$ to $0.7697$ on Movie and increasing from $0.7756$ to $0.7914$ on SpaceX. These results show that Feature Optimization consistently improves predicted diffusion
influence, while its effect on semantic preservation exhibits no clear trend across datasets.

\paragraph{w/o Feature Decoding.}
We remove the decoder $\mathcal{D}$, run LLM-Zero-shot
$m$ times, and retain the candidate with the highest $L_U$,
matching our method's number of content-generation calls
while removing feature guidance. This variant causes the largest reduction in Diffusion Gain, reaching
only $7.96\%$ on Movie and $23.84\%$ on SpaceX. Although Consistency
increases to $0.8663$ and $0.8520$, the generated content becomes
more conservative and less influential. This confirms that
Feature Decoding is crucial for translating optimized
representations into audience-specific guidance.

\paragraph{w/o Multi-Start Generation.}
We use a single initialization and generate all candidates
from the resulting decoded feature. Diffusion Gain decreases to $11.57\%$ on Movie and $34.00\%$ on SpaceX, indicating that a single optimum cannot capture heterogeneous audience interests. Despite higher Consistency scores of $0.8170$ and $0.8081$, the resulting content achieves lower predicted diffusion influence. This result highlights the value of multi-start exploration for identifying diverse and influential generation directions.

\subsection{Discussion}
\label{sec:discussion}

\begin{figure}[t]
\centering\includegraphics[width=\linewidth]{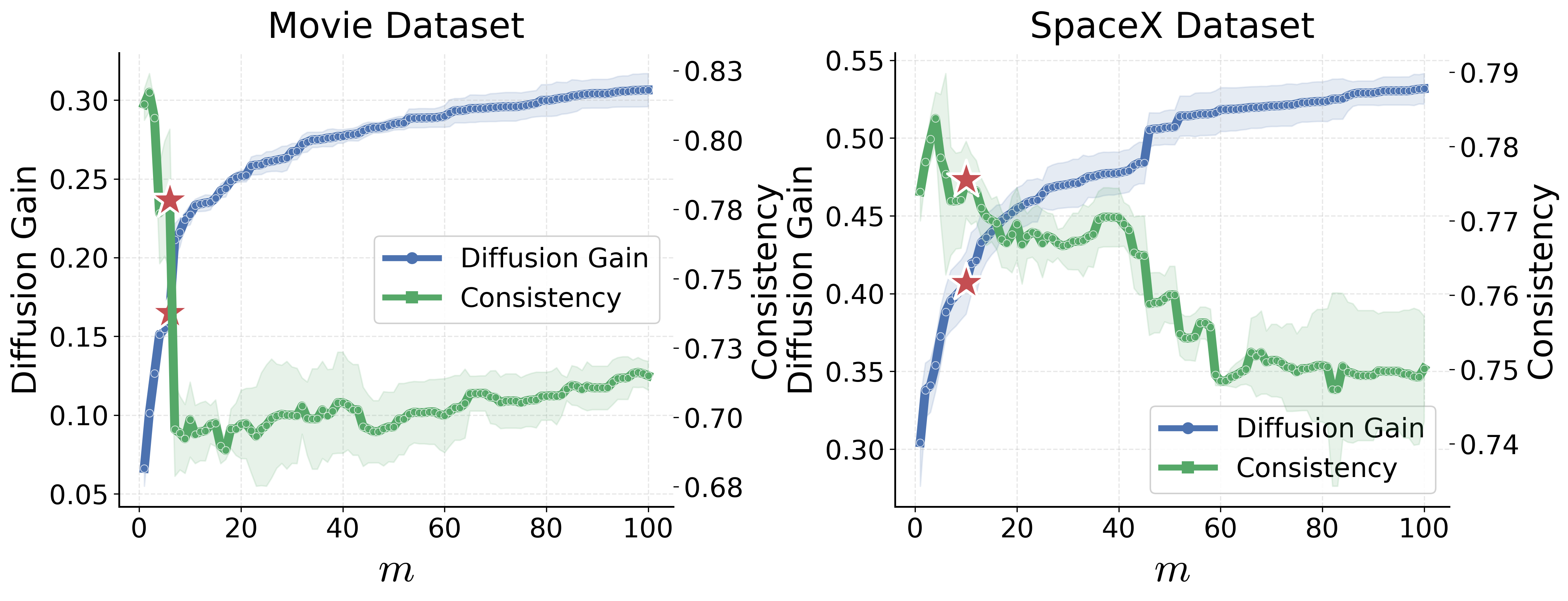}
    \caption{Trade-off between Diffusion Gain and Consistency.}
    \label{fig:m_analysis}
\end{figure}

The performance of our framework is closely linked to the diversity of susceptible features. To determine the optimal number of features $m$, we conduct a sensitivity analysis over $m \in \{1, 2, \dots, 100\}$. As illustrated in Figure~\ref{fig:m_analysis}, Diffusion Gain increases with $m$ before plateauing, while Consistency with the original content generally decreases. This trade-off reflects a tension in diffusion-oriented generation, where improving diffusion may require departing from the source semantic space. Based on the ``elbows'' of the curves, we select $m=6$ for Movie and $m=10$ for SpaceX to balance engagement potential and message integrity.

Beyond this quantitative trade-off, Figure~\ref{fig:qual-pair} illustrates how the framework adapts source content to different target audiences. For the ``Pet Insurance'' example, the Movie version introduces cinematic metaphors such as ``auditioning for an action movie'' and high-drama visual compositions. In contrast, the SpaceX version employs technical and aspirational cues such as ``came through like liftoff,'' accompanied by launch-inspired visual rhetoric. These results demonstrate that our diffusion-oriented approach does not merely apply generic enhancements; instead, it reconfigures content semantics to align with the latent susceptibility patterns of the target audience, thereby facilitating multimodal persuasion.
\section{Conclusion}
\label{sec:conclusion}

We present a three-stage framework for diffusion-oriented
content generation via audience-susceptible feature discovery.
For a target audience group, the framework optimizes
propagation-aware latent features on the realistic content
manifold, decodes them into structured natural-language
descriptions, and uses these
descriptions to guide multimodal content generation conditioned
on the source content. Experiments on two real-world social
media datasets show substantial gains in predicted diffusion
influence, while ablation studies validate the contribution of each module. A user study further shows that participants prefer our
generated content over that of the strongest baseline. Together,
these results demonstrate the value of translating latent influence
signals into interpretable and actionable guidance for
audience-aware content rewriting. Nevertheless, applications
in sensitive domains require stronger safeguards against
misleading or harmful content. Future work will further
investigate robustness, fairness, safety, and human oversight.


\bibliography{aaai2027}


\end{document}